\documentstyle[twoside,fleqn,espcrc2]{article}

\newcommand{\beq}{\begin{equation}}
\newcommand{\eeq}{\end{equation}}

\title{Approximate actions for dynamical fermions}

\author{Alan C. Irving\address{Theoretical Physics Division,
         Department of Mathematical Sciences,
         University of Liverpool,
         PO Box 147, Liverpool L69 3BX, UK.},
   	James C. Sexton
	and Eamonn Cahill\address{
	School of Mathematics,
         Trinity College,
         Dublin 2,
         Ireland
	{\sl and} Hitachi Dublin Laboratory, Dublin 2, Ireland.}\\
	{\em UKQCD Collaboration}
}

\begin{document}

\begin{abstract}
Recent developments and applications of approximate actions for full
lattice QCD are described. We
present first results based on the stochastic
estimation of the fermion determinant on $12^3\times 24$
configurations at $\beta=5.2$.
\end{abstract}

\maketitle

\section{BASIC PRINCIPLES}
\label{sec:princ}

Consider actions $S_1[U]$ and $S_2[U]$ describing two lattice gauge 
theories with the same gauge configuration space $\{ U\}$
so that $(i=1,2)$
\begin{equation}
Z_i\equiv\int {\mathcal D}Ue^{-S_i[U]},\quad
<F>_i\equiv {1\over Z_i}\int {\mathcal D}U e^{-S_i}.
\label{eq:ZF}
\end{equation}
For example, $S_1$ might be
the quenched Wilson action and $S_2$ the SW-improved action
for 2-flavour QCD. Here, $F$ is some operator. 
Expectation values in the two theories can be related
via a cumulant expansion whose leading behaviour implies\cite{ACIJCS} 
\begin{eqnarray}
<F>_2 = <F>_1 +<\tilde{F}\tilde{\Delta}_{12}>_1 + \dots\\
\hbox{where}\quad\Delta_{12} \equiv S_1-S_2,\quad
\tilde{F}\equiv F - <F>\, \hbox{etc.}
\label{eq:F2F1}
\end{eqnarray}

In general, an action is a function of several parameters. e.g.
the Wilson action depends on the bare parameters $\beta$ and $\kappa$. 
One may consider matching different actions in one of several ways 
\cite{ACIJCS}: 
\begin{itemize}
\item[M1:] minimise the \lq distance\rq{} between
the actions, i.e. $\sigma^2(\tilde{\Delta}_{12})$;
\item[M2:] match a given set of operators. i.e. require
$<F_n>_1=<F_n>_2$;
\item[M3:] maximise the acceptance in an exact
algorithm for  $S_2$ constructed via accept/reject applied to
configurations generated with action $S_1$.
\end{itemize}
It turns out that, to lowest order, these 3 tuning prescriptions
coincide. They differ in a calculable way at next order. Details are
in~\cite{ACIJCS}.

\section{APPLICATIONS TO QCD}
\label{sec:applic}

\subsection{Tuning action parameters}
\label{ssec:tune}
In \cite{ACIJCS}, we demonstrated how a relatively modest number of Wilson loop
operators (three or four) 
\beq
S_1 = -\sum_{i=1}^n a_i \tilde{W}_i
\label{eq:Sapprox}
\eeq
can be used to approximate the full Wilson action
\beq
S_2 = S_G - T\quad\hbox{where}\quad 
T\equiv {{n_f}\over{2}}\hbox{TrLn}[M^{\dagger}M]\, .
\label{eq:Sfull}
\eeq

A particularly simple application is to  study the quenched
approximation (see also \cite{SW1}). Here $n=1$ and 
$W_1\approx S_G$, the usual Wilson gauge action. 
Tuning the quenched  coupling
$\beta_Q$ in $S_1$ to match $S_2$ (full QCD) at a given $\beta$ and
$\kappa$ yields (at first order):
\beq
\delta\beta\equiv \beta-\beta_Q= -
{{<\tilde{T}\tilde{W}_1>}\over{<\tilde{W}_1^2>}}\, .
\label{eq:dbeta}
\eeq

In general, one requires a better range of loop operators
\cite{ACIJCS} or, indeed, other types of operators to capture the essential
physics. In all cases, the
basic principles and techniques are the same.
  
\subsection{Bare parameter dependence in QCD}
\label{ssec:bare}
Another practical application is to take
\beq
S_1 = S_{QCD}(\beta_0,\kappa_0)\quad\hbox{and}\quad 
S_2 = S_{QCD}(\beta,\kappa)
\label{eq:Sparam}
\eeq
so as to explore the bare parameter dependence of the lattice theory
using configurations generated at a finite number of
reference points $(\beta_0,\kappa_0)$ in parameter space.
For example, at fixed $\beta=\beta_0$, one might wish to explore the
$\kappa-$dependence of measurements.
According to eqn.~\ref{eq:F2F1}, one requires measurements of 
\beq
\Delta_{12}\equiv S_1 - S_2 = T(\kappa)-T(\kappa_0)\, . 
\label{eq:kshift}
\eeq
One can make stochastic
estimates of this quantity quite efficiently (see section \ref{sec:tech}) - certainly much more
rapidly than by performing independent dynamical fermion simulations
at each set of parameters.

\subsection{Exact algorithm for full QCD}
\label{ssec:exact}
In \cite{ACIJCS}, we constructed an algorithm
which delivers configurations correctly distributed 
with respect to a desired action $S_2$
as follows:
\beq
{\mathcal W}_2(U,U') \equiv 
{\mathcal W}_1(U,U')A(U',\Delta_{12}[U'])\, .
\label{eq:exact}
\eeq
The acceptance probability $A$
depends directly on the quality of the approximate action~\cite{ACIJCS}:

\begin{equation}
A(\Delta_{12})=\hbox{erfc}(\frac{1}{2}\sqrt{\sigma^2(\Delta_{12})})\, .
\label{eq:erfc}
\end{equation}
The action difference $\Delta_{12}$ is of course 
an extensive quantity. In a typical application, 
the transition probability ${\mathcal W}_1$ for the approximate 
action should correspond to a relatively fast update scheme.
Application of this idea to full QCD
may offer a partial solution to decorrelation problems
with standard dynamical fermion algorithms.

\section{STOCHASTIC ESTIMATES OF THE FERMION DETERMINANT}
\label{sec:tech}

We require an unbiased estimator for $\hbox{TrLn}H$ where $H=M^\dagger
M$ is a hermitian positive-definite matrix. Bai, Fahey and Golub
\cite{BFG} have recently proposed  estimators, with bounds,
for quantities of the form
\beq
u^\dagger f(H)v
\label{eq:ufv}
\eeq
where, for example in our case, $f\equiv \hbox{Ln}$.
Taking $u=v$ as some normalised noise vector (e.g. $Z_2$), 
we can obtain a stochastic estimate of $\hbox{TrLn}H$. 
The efficiency of the method, in comparison with the Chebychev-based
methods used previously \cite{SW1,ACIJCS} results from
an elegant relationship between the nodes/weights required for 
a $N$-point Gaussian quadrature and the eigenvalues/eigenvectors
of a Lanczos matrix of dimension $N$ \cite{BFG}.
This relationship and resulting accuracy appears to remain good, 
when orthogonality is lost in standard numerical Lanczos methods.
In the present case, for a fixed noise vector we obtain 6 
figure convergence of the quadrature with 70 Lanczos steps
on a matrix with condition number of order $10^4$.
The estimator of $T\equiv \hbox{TrLn}H$ is of the form
\beq
E_T={1\over{N_{\phi}}}\sum_{i=1}^{N_\phi} I(\phi_i)\, ,\quad
I(\phi_i) = \sum_{j=1}^N\omega_j\hbox{Ln}(\lambda_j)
\label{eq:ET}
\eeq
where $\{\lambda_j\}$ are eigenvalues of the tridiagonal
Lanczos matrix arising using $\phi_i$ as a starting vector and
$\{\omega_j\}$ are related to the corresponding eigenvectors. 
Details of the method will be presented elsewhere \cite{CIS}. 

\section{QCD RESULTS AT $\mathbf{\beta=5.2}$}
\label{sec:res}

\subsection{Quenched comparisons}
\label{ssec:quen}
In Table~\ref{tab:bshift} we show 
the results of matching quenched and SW-improved \cite{SW} 
fermion actions using eqns.~\ref{eq:F2F1},~\ref{eq:dbeta}.
\begin{table}[ht]
\caption{\it Equivalent quenched $\beta$ results predicted
from full QCD data at $\beta=5.2$.
$\beta_Q$ is the equivalent coupling,
$<P>$ is the measured average plaquette, and $<P>_Q$
the actual quenched value at $\beta_Q$.
}
\label{tab:bshift}
\centering
\begin{tabular}{ccccc}
\hline
        $\kappa$    &Configs.  &$<P>$	&$\beta_Q$  &$<P>_Q$ \\ 
\hline
        $.136$ &$100$   	&$.4874(1)$	&$5.48(2)$	&$.488(4)$ \\
        $.139$ &$\phantom{0}40$    &$.5160(2)$	&$5.62(2)$	&$.530(4)$ \\
\hline
\end{tabular}
\end{table}
The errors on comparison values of $<P>_Q$ include 
those associated with the uncertainty in estimating $\beta_Q$. 
The comparison at the heavier quark mass, $\kappa=.136$, shows
complete consistency ($\approx 0.488$). 
That at the lighter quark mass is close but 
not precise. 
It is seen from this and from other results that,
at $\kappa=.139$, one may be starting to see non-trivial
effects of un-quenching. 

\subsection{Static potential measurements}
\label{ssec:pot}
To probe dynamical quark effects more closely, we have studied
the static potential on quenched ($\beta_Q=5.7$) and on 2 flavour
dynamical fermions configurations ($\beta=5.2$) at several values of $\kappa$. 
Measurements were made using the optimised techniques of 
\cite{CMpot}. Results for
$V(R)$ and the extraction of the lattice spacing
via $r_0$~\cite{R0} are presented elsewhere \cite{MTalevi}.
Strong scale dependence on $\kappa$ is observed. 
We used eqns. \ref{eq:F2F1} and \ref{eq:kshift}
(working to lowest order) to \lq predict\rq{} the 
$\kappa=.139$ potential from that measured at $\kappa=.136$ where the
quarks are effectively much heavier.
There was evidence that the lowest order correction captures
much of the change in behaviour but with large errors.
This is not surprising since making comparisons at fixed $\beta$
involves different physical scales.

\subsection{Estimating \lq distances\rq{}}
\label{ssec:dist}
In Table~\ref{tab:dist}, we show the distance (squared)
$\sigma^2(\Delta_{12})$,
as defined in section \ref{sec:princ},
between our dynamical fermion reference action at $\beta_0=5.2$, $\kappa_0=.136$ and
other relevant actions. 
\begin{table}[ht]
\caption{\it Estimated distance squared $\sigma^2(\Delta_{12})$.
}
\label{tab:dist}
\centering
\begin{tabular}{cccc}
\hline
        $Action$    	&$\beta$  &$\kappa$	&\lq Distance\rq\\
\hline
        Quen. 	&$5.2\phantom{00(00)}$   	&$-$	&$9250(1330)$	\\
        Quen. 	&$5.479(15)$   	&$-$	&$855(255)\phantom{00}$	\\
        $N_f=2$	&$5.2\phantom{00(00)}$		&$.139$	&$147(20)\phantom{000}$	\\
        $N_f=2$	&$5.165(1)\phantom{0}$	&$.139$	&$17(3)\phantom{00000}$\\
\hline
\end{tabular}
\end{table}
For example, the second line shows the quality of the quenched tuning
described above (see Table~\ref{tab:bshift}): a reduction from 9250 to
855 by shifting $\beta$. 
Also, we note that $\kappa=.139$ is some distance
($147$) from the reference $\kappa_0=.136$. This is the origin of the
difficulty (noted above) in predicting $\kappa=.139$ results 
from $.136$ configurations
{\em without also tuning} $\beta$. 
The last line of the table shows that
operator matching should be much closer using 
$(\beta_0,\kappa_0)=(5.2,0.136)$ 
and 
$(\beta,\kappa)=(5.165,0.139)$, since the estimated 
distance is only $17\pm3$.
This could be checked directly with further simulations at 
$\beta=5.165$, $\kappa=.139$. We would expect the lattice spacing to be
similar at these two points in action parameter space.

\subsection{Approximate loops actions}
\label{ssec:loops}

We have begun a systematic study of possible approximate actions 
constructed from Wilson loops. Initially, we have chosen to construct
these from a range of link steps of size $1,2,3,4$ etc. 
The aim is to minimise the distance between the target
action, e.g. full QCD at $\beta_0=5.2$ and $\kappa_0=.136$ in the 
present example, and the parametrised loop action \cite{ACIJCS}.
It is not hard to reduce the distance well below that achieved
with a single plaquette (see Table~\ref{tab:dist}). 
However, eqn.~\ref{eq:erfc} shows, for eample, that $\sigma^2$ must be less than 10 
to give an acceptance greater than 2.5\%, in the
corresponding exact algorithm.

Further work will include:
continuing studies to establish more efficient loop operators
and other classes of approximate action; 
studies of the parameter space of
standard dynamical fermion actions; 
mapping the parameter spaces of different types of action.

\end{document}